\documentclass[prb,twocolumn,showpacs,preprintnumbers,amsmath,amssymb]{revtex4}

\usepackage{amsmath,amssymb,amsfonts}
\usepackage[dvips]{graphicx}
\usepackage[latin1]{inputenc}
\usepackage{subfigure}

\begin{document}

\title{Accurate Ground State Energies of Solids and Molecules from Time Dependent Density
  Functional Theory}

\author{Thomas Olsen}
\email{tolsen@fysik.dtu.dk}
\author{Kristian S. Thygesen}

\affiliation{Center for Atomic-Scale Materials Design (CAMD) and Center for Nanostructured Graphene (CNG),
	     Department of Physics, Technical University of Denmark,
	     DK--2800 Kongens Lyngby, Denmark}

\date{\today}

\begin{abstract}
  We demonstrate that ground state energies approaching chemical accuracy
  can be obtained by combining the adiabatic connection
  fluctuation-dissipation theorem (ACFDT) with time-dependent density
  functional theory (TDDFT). The key ingredient is a renormalization scheme, which
  eliminates the divergence of the correlation hole characteristic of any local kernel. 
This new class of renormalized kernels gives a significantly better
  description of the short-range correlations in covalent bonds
  compared to the random phase approximation (RPA) and yields a four fold
  improvement of RPA binding energies in both molecules and solids. We
  also consider examples of barrier heights in chemical reactions,
  molecular adsorption and graphene interacting with metal surfaces,
  which are three examples where RPA has been successful. In these
  cases, the renormalized kernel provides results that are of equal quality
  or even slightly better than RPA, with a similar computational cost.
\end{abstract}

\maketitle
The calculation of ground state energies with a precision on the order of k$_B$T at room temperature (around 1 kcal/mol or 43 meV) is a long standing challenge in computational chemistry and materials science. This degree of precision, sometimes referred to as "chemical accuracy", can in principle be achieved using quantum chemistry wave function methods \cite{bartlett}. However, the extreme scaling of such methods limits their usage to relatively small systems, and their application to periodic systems in general and metals in particular, seems highly non-trivial \cite{alavi, shepherd}.

Recently, the adiabatic-connection fluctuation-dissipation theorem (ACFDT) has become a popular method for obtaining electronic ground state energies form first principles. With this approach the electronic correlation energy is obtained from an approximation to the dynamic density response function $\chi^\lambda(\omega)$, which can be obtained from Time-Dependent Density Functional Theory (TDDFT) through the Dyson equation
\begin{equation}\label{dyson}
\chi^\lambda(\omega)=\chi^{KS}(\omega)+\chi^{KS}(\omega)f^\lambda_{Hxc}(\omega)\chi^\lambda(\omega).
\end{equation}
Here $\chi^{KS}(\omega)$ is the non-interacting Kohn-Sham response function and $f^\lambda_{Hxc}(\omega)=\lambda v_c+f^\lambda_{xc}(\omega)$ with $v_c$ the Coulomb interaction and $f_{xc}^\lambda(\omega)$ the xc-kernel. In static Density Functional Theory (DFT), one needs a rather involved approximation for the xc-energy in order to provide a decent description of the ground state properties of a particular system and no approximation has so far been able to provide accurate results across different binding regimes. In contrast, the ACFDT and Eq. \eqref{dyson} gives a simple framework for calculating correlation energies in terms of the excited states of a non-interacting auxiliary system and is expected to provide a high degree of accuracy with simple approximations for the xc-kernel. In particular, if we use $f_{xc}=0$ we obtain the Random Phase Approximation (RPA) \cite{bohm1}, which has been shown to give
 an accurate account of dispersive interactions, static correlation, weak covalent bonds, and is presently considered state of the art in \textit{ab initio} electronic structure theory involving solid state systems \cite{furche,eshuis,ren_review,olsen_rpa1,olsen_rpa2,jun_rpa,harl09,harl10,schimka}. Nevertheless, RPA suffers from large self-correlation errors and predicts too weak binding of solids and molecules, which have severely limited the universal applicability of the method. Furthermore, extending RPA with semi-local approximations for the xc-kernel has been shown to fail dramatically, and progress in "beyond RPA" methods within the framework of TDDFT has so far been very limited \cite{lein,furche_voorhis,dobson_wang00,jung,gould,hesselmann1,hesselmann2}. A somewhat orthogonal approach for improving RPA ground state energies, is based on eliminating the RPA self-correlation energy within many-body perturbation theory and is referred to as second-order screened exchange (SOSEX). The SOSEX correlation energy vanishes by construction for any one-electron system and has been shown to improve the accuracy of covalent bonds slightly, while it completely deteriorates the good description of static correlation and barrier heights in RPA \cite{gruneis, ren_review}. In addition, SOSEX scales as $N^5$ with system size and therefore quickly becomes much more computationally demanding than RPA, which scales as $N^4$.

\begin{figure}[tb]
\begin{center}
 \includegraphics[scale=0.40]{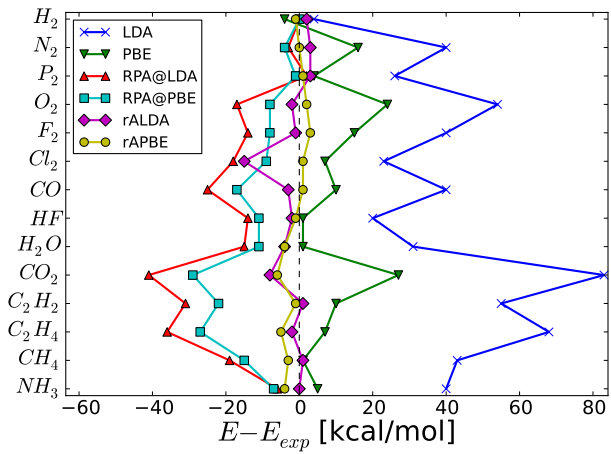}
 \caption{(Color online) Molecular atomization energies evaluated with different methods shown relative to the experimental values. Results are shown with respect to reference values from Ref. \cite{karton}. The numbers are tabulated in the supplementary material.}
\label{fig:energies_mol}
\end{center}
\end{figure} 
In this Letter, we apply a renormalization scheme that solves the divergence problem associated with semi-local adiabatic TDDFT in the ACDFT. We have implemented the xc-kernels for the two most commonly used xc-approximations (LDA and PBE) \cite{pbe} within this renormalization scheme and show that the results obtained with adiabatic renormalized versions are superior to RPA in all cases. In particular, with the renormalized adiabatic PBE we obtain four-fold improvement of binding energies of molecules and solids, while it maintains the $N^4$ scaling of RPA.

The construction of the renormalized kernel is motivated by the observation that the Fourier transformed correlation hole in the homogeneous electron gas (HEG) is closely reproduced by the ALDA exchange kernel for wave vectors $q<2k_F$. For larger $q$ the exact hole is essentially zero. In contrast, while the ALDA hole vanishes exactly for $q=2k_F$, it attains a finite value for larger $q$ and decays to zero much too slowly. In fact, this long tail produces a divergence of the correlation hole at $r=0$. For $q=2k_F$ the ALDA correlation hole vanishes, since $f_{Hxc}^{ALDA}(2k_F)=4\pi/(2k_F)^2+f_{x}^{ALDA}[n]=0$, and it is tempting to define a new renormalized kernel as $f_{Hxc}^{rALDA}(q)=\theta(2k_F-q) f_{Hxc}^{ALDA}(q)$ (note that in this work we only consider the exchange part of the LDA and PBE kernels). Alternatively, one can generalize the LDA energy functional to include non-local exchange-correlation effects, by replacing the local density $n(\mathbf{r})$ by an averaged quantity $n^*(\mathbf{r})=\int\phi(\mathbf{r}-\mathbf{r}')n(\mathbf{r}')d\mathbf{r}'$. This substitution will not alter the ground state energy or potential of the homogeneous electron gas, but it will introduce non-locality into the kernel. In fact, it seems physically reasonable that the exchange-correlation energy density should include contributions from the density in the vicinity of the xc hole, which will be the case if $\phi$ has a width similar to the xc hole. This is accomplished by choosing $\phi$ as the Fourier transform of the step function $\theta(2k_F-q)$ and we obtain the renormalized kernel as the adiabatic kernel derived from the generalized LDA energy functional. In the supplementary material we provide more details on this derivation.  

In Refs. \cite{olsen_ralda1,olsen_ralda2} we investigated this rALDA kernel in detail and we refer to those papers for more details. The idea is readily generalized to any adiabatic semi-local kernel, AX, and we obtain the following expression for the renormalized kernel in real space:
\begin{align}\label{rX}
f_{xc}^{rAX}[n](r)&=\frac{f^{AX}_{xc}[n]}{2\pi^2r^3}\Big[\sin(q_c[n]r)-q_c[n]r\cos(q_c[n]r)\Big],\notag\\
&\qquad-\frac{1}{r}\Big[1-\frac{2}{\pi}\int_0^{q_c[n]r}\frac{\sin x}{x}dx\Big],
\end{align}
where
\begin{align}\label{qc}
q_c[n]=\sqrt{\frac{-4\pi}{f^{AX}_{xc}[n]}}.
\end{align}
Here $q_c[n]$ is the cut-off wave vector where the Fourier transform of the correlation hole of the HEG obtained with $f^{AX}_{xc}$ becomes zero (for X=LDA the cut-off becomes $q_c=2k_F$). To generalize the construction to inhomogeneous systems we apply the substitutions
\begin{align}\label{inhomo}
r&\rightarrow|\mathbf{r}-\mathbf{r'}|\\
n&\rightarrow\frac{n(\mathbf{r})+n(\mathbf{r'})}{2}\\
\nabla n&\rightarrow\frac{\nabla_\mathbf{r}n(\mathbf{r})+\nabla_\mathbf{r'}n(\mathbf{r'})}{2}.
\end{align}
Thus both $f_{xc}^{rAX}$ and $q_c$ become functions of $\mathbf r$ and $\mathbf r'$. Importantly, the delta function of the original adiabatic kernel acquires a finite (density-dependent) width which ensures a finite value of the correlation hole for $\mathbf r = \mathbf r'$ and entails a better description of the short-range correlation effects. We will only discuss non-self-consistent applications of the kernel. This implies that calculated energies will depend on the input density, orbitals and eigenvalues. This situation is well known from RPA calculations. We note, however, that in contrast to RPA where no obvious starting point exists, it is natural to base an rAX calculation on a DFT-X calculation. We regard this is as a fundamental merit of the method, since it eliminates the arbitrary choice of input orbitals in non-self-consistent RPA calculations. 
\begin{figure}[tb]
\begin{center}
 \includegraphics[scale=0.3]{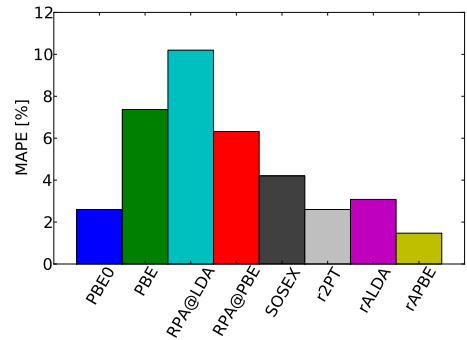}
 \caption{(Color online) Mean absolute percentage deviation of molecular atomization energies. The PBE0, SOSEX, and rP2T values are taken from Ref. \cite{ren_review}.}
\label{fig:mape_mol}
\end{center}
\end{figure} 

We have previously shown that the class of kernels \eqref{rX} significantly improves the correlation energies of the Homogeneous electron gas. Furthermore, using the exchange part of rALDA, reduces the correlation energy of a Hydrogen atom to -0.1 eV. This is already a major improvement compared to RPA which gives $E_c=-0.6$ eV, but the value still comprises a significant self-correlation error. However, inclusion of gradient corrections in the form of PBE exchange, reduces the correlation energy to less than one meV (less than the resolution of the implementation). In the following we will refer to this kernel as rAPBE and demonstrate the superiority of the resulting kernel over the RPA and other "beyond RPA" methods \footnote{For simplicity, we have neglected the variation in the potential with respect to the density gradient and thus use $f^{APBE}_x[n]=\frac{\partial v_x^{PBE}(n)}{\partial n}\Big|_{n=n(\mathbf{r})}$ to construct the rAPBE kernel.}. The method has been implemented in a plane wave basis in the electronic structure GPAW \cite{gpaw-paper, jun, olsen_rpa2, olsen_ralda2}, which uses the PAW method \cite{blochl}. We refer to the supplementary material for computational details. We note here that the evaluation of RPA correlation energies is dominated by the calculation of $\chi^0$, which scales as $N^4$. Evaluating the kernel \eqref{rX} scales as $N^2$ and the method therefore scales as RPA with system size.

\begin{figure}[tb]
\begin{center}
 \includegraphics[scale=0.4]{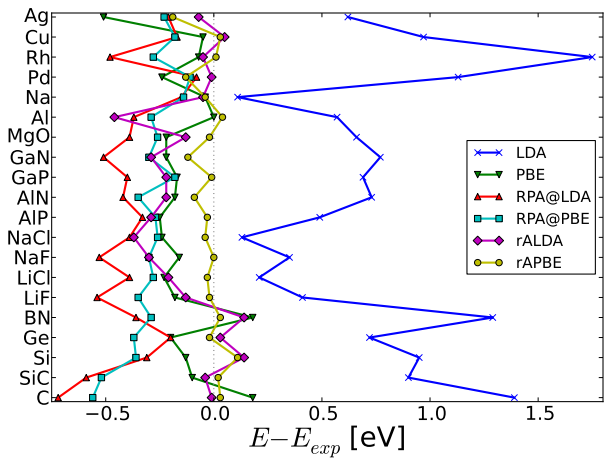}
 \caption{(Color online) Deviation from experimental values of the cohesive energy of solids evaluated with different functionals. The numbers are tabulated in the supplementary material.}
\label{fig:energies_sol}
\end{center}
\end{figure} 
Fig. \ref{fig:energies_mol} shows the deviation from experimental values of molecular atomization energies. RPA has a well-known tendency to underbind and performs somewhat worse than PBE. We also observe a significant difference between RPA@LDA and RPA@PBE with RPA@PBE being the more accurate. The renormalized kernels rALDA and rAPBE shows a striking improvement compared to both RPA and PBE with most errors being on the order of 1-3 kcal/mol. In Fig. \ref{fig:mape_mol} we compare the mean absolute relative errors of different methods. It is seen that already the rALDA is better than the SOSEX and approaches the accuracy of r2PT (SOSEX corrected for single excitations \cite{ren}) and the hybrid PBE0. However, the gradient-corrected rAPBE reduces the error of rALDA by a factor of two and outperforms both PBE0 and rPT2. Since RPA already performs very well for geometry optimization, we have not performed a detailed comparison of bond lengths with the renormalized kernels. We have only considered the F$_2$ and N$_2$ molecules where we find deviations from the experimental bond lengths of $1.6\%(2.4\%)$ and $0.4\%(1.2\%)$ with rAPBE (RPA), respectively.

In contrast to the case of atomization energies, RPA is highly accurate for molecular barrier heights where it performs much better than both the SOSEX methods and hybrid functionals \cite{ren_review}. Here we have just tested a single reaction barrier with the rAPBE functional namely: H+N$_2$O$\rightarrow$OH+N$_2$. For this case we obtain errors of 0.9(0.4) kcal/mol and  0.2(2.4) kcal/mol for the forward and backward reactions, respectively, using the rAPBE(RPA) functional. For comparison we obtain errors of 3.2(5.9) and 12.2(8.7) using PBE0(B3LYP). Another important property of the renormalized kernels, is the fact that they describe the dissociation of the H$_2$ molecule correctly in the strict atomic limit \cite{olsen_ralda2}. This is in contrast to most MBPT extensions of RPA, where one has to sacrifice the good description of static correlation in molecules \cite{ren_review}.

\begin{figure}[tb]
\begin{center}
 \includegraphics[scale=0.3]{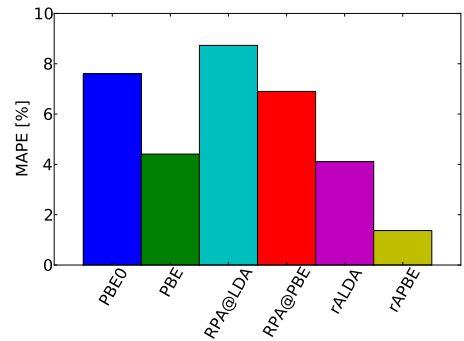}
 \caption{(Color online) Mean absolute percentage deviation of the cohesive energy of solids The PBE0 results are from Ref. \cite{paier}. }
\label{fig:mape_sol}
\end{center}
\end{figure}

While the hybrid functionals may be a good choice for a decent accuracy in molecular atomization energies, these methods fail completely for the cohesive energies of solids. The RPA also performs poorly, and the PBE functional seems to be the best choice for this problem. Nevertheless, from Figs. \ref{fig:energies_sol} and \ref{fig:mape_sol} it is clear that the rALDA functional outperforms RPA and produces an accuracy similar to PBE. Inclusion of gradient corrections through the rAPBE functional reduces the error by a factor of four and is thus much more accurate than any of the other functionals considered. We note in passing that SOSEX has been shown to produce results of similar accuracy as the rAPBE for a small test set of five semiconductors \cite{gruneis}, but scales as $N^5$ with system size and therefore becomes significantly more computationally demanding for solid state systems than RPA. We have confirmed that the rAPBE functional inherits the good description of lattice constants within RPA \cite{harl10}. In particular, we have calculated the lattice constants for bulk C, Si, Na and Pd. The results are provided in the supplementary. The deviations from experimental values corrected for zero point anharmonic effects are $0.4\%(0.8\%)$, $0.6\%(0.6\%)$, $1.8\%(1.4\%)$, 
and $1.8\%(1.4\%)$ for rAPBE(RPA) respectively.

One might speculate that the rAPBE accuracy of atomization energies and cohesive energies of solids are simply due to a better description of the isolated atoms compared to RPA. Of course, the already much improved correlation energy of the HEG compared to RPA indicates that this is not the case. As another demonstration of this fact we have calculated the formation energy of MgO defined as the energy of the reaction Mg(s) + 1/2O$_2$(g) $\rightarrow$ MgO(s). The result is shown in Tab. \ref{tab:formation}. As can be seen, rAPBE reduces the RPA error by $0.1$ eV corresponding to a four-fold decrease in relative error. Thus, the renormalized kernel does not only improve the description of isolated atoms relative to RPA, but gives rise to a universal improvement of correlation energies that manifests itself in any calculated quantity.
\begin{table}[b]
\begin{center}
\begin{tabular}{c|c|c|c|c}
 PBE & HF & RPA & rAPBE & Exp\\
	\hline
5.34  & 6.04 & 6.12 & 6.22  & 6.26
\end{tabular}
\end{center}
\caption{Formation energies of MgO calculated with different functionals. All numbers are in eV. The experimental value has been corrected for zero-point energy \cite{harl09}.}
\label{tab:formation}
\end{table}

\begin{figure}[tb]
\begin{center}
 \includegraphics[scale=0.40]{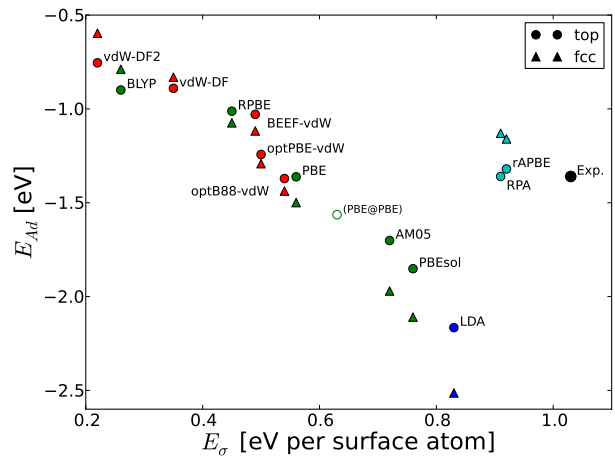}
 \caption{(Color online) Surface energy vs adsorption energy of CO/Pt(111) calculated with various GGA functionals (green markers) and van der Waals functionals (red markers). Circles and triangles indicate atop and hollow sites, respectively. All calculations were performed with the experimental lattice constant of Pt and the CO molecule relaxed with PBE. The hollow circle was obtained with a PBE optimized lattice constant.}
\label{fig:surf}
\end{center}
\end{figure}

One of the great success stories of RPA, is that it solves the famous "CO puzzle". Most GGA functionals predict the wrong adsorption site for CO on metal surfaces, whereas RPA predict the correct adsorption order \cite{schimka}. Furthermore, it is possible to choose a semi-local functional that gives the correct adsorption energy, but this will be at the cost of a highly inaccurate metal surface energy. In contrast, RPA produces an accurate adsorption energy and still yields a reasonable metal surface energy. As shown in Fig. \ref{fig:surf}, this trend is inherited by the rAPBE functional which produce results very similar to those of RPA. For comparison we also show the results for various GGAs \cite{schimka} and van der Waals functionals \cite{wellendorff}.

\begin{figure}[tb]
\begin{center}
 \includegraphics[scale=0.40]{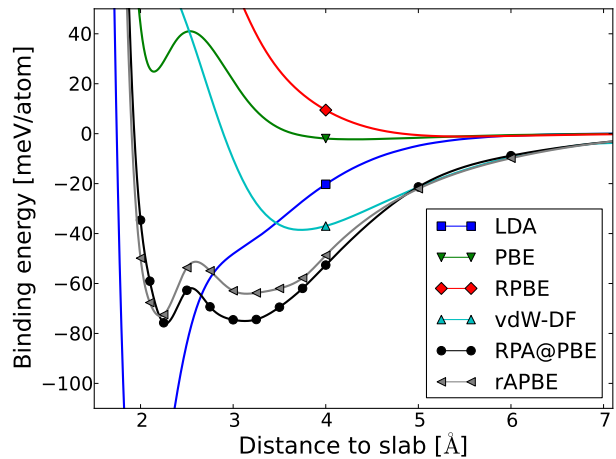}
 \caption{(Color online) Potential energy curve for graphene on Ni(111) calculated with different methods. The experimental binding distance is $d\approx 2.1$~\AA.}
\label{fig:graph}
\end{center}
\end{figure} 
Another successful application of the RPA method is the adsorption of graphene on metal surfaces \cite{olsen_rpa1, mittendorfer, olsen_rpa2}. Graphene interacts with metals through long range dispersive- and short range weak covalent interactions. The adsorption geometry and binding energy is determined by a detailed balance between these contributions which are of equal magnitude. Semi-local functionals cannot account for the dispersive interactions and in many cases do not provide the required accuracy for weak covalent interactions either. On the other hand, most van der Waals functionals account well for the dispersive interactions at large distances, but fail at shorter distances due to incorrect description of the exchange interaction\cite{wellendorff, mittendorfer}. The RPA predict metal-graphene binding distances in overall good agreement with experiments. In particular, for the case of Ni(111) and Co(0001) RPA yields two distinct minima around $d=2.2$ {\AA} and $d=3.3$ {\AA} corresponding to (weak) chemisorption and physisorption, respectively. The chemisorption minimum is slightly deeper in good agreement with experimental findings of $d\approx 2.1$ {\AA}. However, due to the inaccurate description of covalent bonds in molecules and solids one could question the accuracy of the RPA description of the chemisorption minimum. So far, RPA has been the best possible method to analyze these systems, since quantum chemistry methods are out of the question for this class of systems. We have calculated the binding energy curve for graphene on Ni(111) using the rAPBE kernel and compared it to various other methods, see Fig. \ref{fig:graph}. It is seen that rAPBE and RPA produce very similar results. As one could perhaps expect from the general tendency of RPA to underbind, the rAPBE kernel lowers the chemisorption minimum relative to the physisorption minimum. While this change is in the right direction compared to experiments, it does not change the qualitative picture of the binding obtained from RPA.

In conclusion, we have shown how to construct xc-kernels within TDDFT
that extends the RPA in a natural way and improves its description of
short-range correlations while retaining the good description of
dispersive interactions and static correlation with similar
computational cost. The proposed renormalization procedure can be
applied to any known semi-local xc-functional and thus defines an
entirely new class of adiabatic non-local xc-kernels which could pave
the way for achieving chemical accuracy in solid state calculations.

The authors acknowledge support from the Danish Council
for Independent Research's Sapere Aude Program, Grant
No. 11-1051390. The Center for Nanostructured Graphene
is sponsored by the Danish National Research Foundation,
Project DNRF58.

%\bibliography{bibfile}{}

\newpage

\appendix
\section{The renormalized kernel from a weighted density approximation}
Here we will show how the renormalized kernels arise naturally as a generalization of (semi-)local functionals, if one replaces the local density by a modified density obtained as an average of the original density over the extend of the xc-hole. In the case of the homogeneous electron gas, the LDA exchange-correlation functional gives the exact ground state energy (by construction). However, the adiabatic kernel obtained from LDA gives rise to poor correlation energies and has the unphysical property that the pair-correlation function diverges at the origin. Nevertheless, it has previously been shown that the main problem with the exchange-correlation kernel is not the lack of frequency dependence, but rather the local nature of the exchange-correlation kernel. Thus, the adiabatic approximation in itself does not seem to pose a fundamental problem for ground state energies within the ACDFT framework. It is therefore natural to ask whether it is possible to generalize LDA such that it gives a more consistent description of exchange-correlation in the sense that both the energy functional itself and the energy obtained from the ACDFT using the derived adiabatic kernel yield accurate ground state energies of the HEG.

To that end, we consider replacing the density in $E_{Hxc}[n]$ by a locally averaged density of the form
\begin{align}\label{n_average}
n^*(\mathbf{r})=\int d\mathbf{r}'\phi(\mathbf{r}-\mathbf{r'})n(\mathbf{r'}),
\end{align}
where $\phi$ is a suitably chosen weighting function. On physical grounds, $\phi$ should integrate to 1 and have a width of around $r_s$ which is the characteristic size of the xc-hole in the HEG. Apart from this, $\phi$ could be considered as a free parameter to be optimized such as to yield the best ACDFT correlation energy of the HEG. It is clear that the substitution (1) does nothing for a constant density and thus the LDA remains exact for homogeneous systems. It does, however, affect the xc-kernel which becomes non-local. Specifically we get
\begin{align}\label{f_average}
E_{Hxc}^*[n]=&E_{Hxc}[n^*]\\
v_{Hxc}^*[n](\mathbf{r})=&\frac{\delta E_{Hxc}^*[n]}{\delta n(\mathbf{r})}=\int d\mathbf{r}'v_{Hxc}[n^*](\mathbf{r}')\phi(\mathbf{r}'-\mathbf{r})\notag\\
f_{Hxc}^*[n](\mathbf{r}-\mathbf{r}')=&\frac{\delta v_{Hxc}^*[n](\mathbf{r})}{\delta n(\mathbf{r'})}\notag\\
=\int &d\mathbf{r}''d\mathbf{r}''' f_{Hxc}[n^*](\mathbf{r}''-\mathbf{r}''')\phi(\mathbf{r}''-\mathbf{r})\phi(\mathbf{r}'''-\mathbf{r}').\notag
\end{align}
For inhomogeneous systems the substitution \eqref{n_average} will of course modify the energy and potential functionals, but the HEG represents a fixed point for which the energy and potential remains the same. Therefore, one could argue that an LDA$^*$ based on \eqref{n_average} may be equally justified and that the usual LDA is just a special case corresponding to the choice $\phi(r-r')=\delta(r-r')$. In fact, for an inhomogeneous system it seems physically reasonable to use the energy density of a HEG with the locally averaged density rather than the strict local density because the xc-hole is not local anyway, but extends over $\sim r_s$. Moreover, in the context of the ACFDT formalism, the use of a non-local $\phi$ will produce a non-local kernel, which is known to be a crucial property. 

The double convolution in Eq. \eqref{f_average} becomes a simple product in Fourier space. A particular simple form for the kernel in Eq. \eqref{f_average} is obtained by choosing $\phi$ to be the Fourier transform of a step function of the form $\theta(q_c-|q|)$. To ensure that the renormalized kernel is continuous in $q$, we must fix $q_c$ by the condition $f_{Hxc}[n](q_c)=0$. We note that in the case of pure LDA exchange, this condition becomes $q_c=2k_F$ meaning that $\phi$ will have a characteristic width of $\sim 1/k_F$ corresponding to the size of the xc-hole in the HEG.

The above discussion, in particular the derivation of the renormalized kernel $f^*$, was mainly focused on the HEG. To obtain a kernel appropriate for inhomogeneous systems it is natural to make a local density approximation for the renormalized kernel, i.e.
\begin{align}
f^{inhom}_{Hxc}[n](\mathbf{r},\mathbf{r}') = f^*_{Hxc}[n(\mathbf{r})/2+n(\mathbf{r}')/2](|\mathbf{r}-\mathbf{r}'|).
\end{align}
This is the kernel we have investigated in the present work. It should be noted that we implicitly assumed that $\phi$ was independent of the density when deriving \eqref{f_average}. Nevertheless, the physical contents and the idea of deriving non-local kernels from LDA$^*$ type functionals should be clear from the above discussion.

\section{Implementation}
The response function and exchange-correlation kernel has been implemented in a plane wave basis set. The kernel Eq. (1) is invariant under simultaneous lattice translations in $\mathbf{r}$ and $\mathbf{r}'$ and its plane wave representation takes the form
\begin{align}\label{f_GG}
f^{rX}_{\mathbf{G}\mathbf{G}'}(\mathbf{q})&=\frac{1}{NV}\int_{NV}d\mathbf{r}\int_{NV}d\mathbf{r}'e^{-i(\mathbf{G+q})\cdot\mathbf{r}}\\
&\qquad\qquad\qquad\times f^{rX}(\mathbf{r},\mathbf{r}')e^{i(\mathbf{G'+q})\cdot\mathbf{r}'}\notag\\
&=\frac{1}{V}\int_{V}d\mathbf{r}\int_{V}d\mathbf{r}'e^{-i\mathbf{G}\cdot\mathbf{r}}f(\mathbf{q};\mathbf{r},\mathbf{r}')e^{i\mathbf{G}'\cdot\mathbf{r}'}.\notag
\end{align}
Here $\mathbf{G}$ and $\mathbf{G}'$ are reciprocal lattice vectors, $N$ is the number of sampled unit cells, $\mathbf{q}$ belongs to the first Brillouin zone, and
\begin{align}\label{f_tilde}
f(\mathbf{q};\mathbf{r},\mathbf{r}')=\frac{1}{N}\sum_{i,j}e^{i\mathbf{q}\cdot\mathbf{R}_{ij}}e^{-i\mathbf{q}\cdot(\mathbf{r}-\mathbf{r}')}f^{rALDA}_{x}(\mathbf{r},\mathbf{r}'+\mathbf{R}_{ij}),
\end{align}
where we have introduced the lattice point difference $\mathbf{R}_{ij}=\mathbf{R}_i-\mathbf{R}_j$ and used that each of the $N$ sampled unit cell integrals in Eq. \eqref{f_GG} can be transferred into a single unit cell by letting $\mathbf{r}\rightarrow\mathbf{r}+\mathbf{R}_i$. We also used that $f^{rALDA}_{Hx}(\mathbf{r}+\mathbf{R}_j,\mathbf{r}'+\mathbf{R}_i)=f^{rALDA}_{Hx}(\mathbf{r},\mathbf{r}'+\mathbf{R}_{ij})$. The function $f(\mathbf{q};\mathbf{r},\mathbf{r}')$ is thus periodic in both $\mathbf{r}$ and $\mathbf{r}'$. In principle the sum over lattice point differences should be converged independently, but in all the calculations in the present work we have set the number of sampled unit cells in the kernel equal to the \textit{k}-point sampling.

\subsection{Computational details}
The correlation energy is given by 
\begin{align}\label{E_c}
E_c=-\int_0^1d\lambda\int_0^\infty\frac{d\omega}{2\pi}\text{Tr}[v\chi^\lambda(i\omega)-v\chi^0(i\omega)],
\end{align}
where the interacting response function $\chi^\lambda(i\omega)$ is calculated from the Dyson equation Eq. (1). We have only considered exchange kernels, which have the property $f_x^\lambda=\lambda f_x$. In principle, the method could easily be applied to kernels including correlation, since the coupling constant scaling is well-known, but this would then require an evaluation of the kernel at all sampled $\lambda$-points. The coupling constant integration in Eq. \eqref{E_c} is evaluated numerically using 8 Gauss-Legendre points. The frequency integration is also evaluated numerically using 16 Gauss-Legendre points distributed from 0-800 eV. We have checked that these samplings produce converged results for a wide range of different electronic systems.

In all calculations the total number of bands used to evaluate $\chi^0$ was set equal to the number of plane waves in the calculation. This ensures a $E_{cut}^{-3/2}$ convergence of correlation energies with respect to the plane wave cutoff $E_{cut}$. For all systems we have evaluated the correlation energies by calculating the expression Eq. \eqref{E_c} at a sequence of correlation energies up to 300-400 eV and then extrapolated energy differences to the infinite cutoff limit. This procedure has previously been shown to be highly accurate for RPA and the convergence properties of the rAX are very similar to that of RPA. The initial input orbitals was calculated with a cutoff of 600 eV and the Hartree-Fock energies were all calculated with 800 eV.

For most molecular and atomic systems the correlation energies were computed in a periodic unit cell with at least 6 {\AA} between neighboring cells. For all metal atoms as well as the Cl and P atoms a separation of 8 {\AA} was required and for Na  we used a separation of 10 {\AA} to ensure decoupling of periodic images. The Hartree-Fock and hybrid calculations were converged separately, since much larger unit cells were required for those calculations.

For the bulk solid state systems, we have used \textit{k}-point samplings of 12x12x12 for Ag, Cu, Rh, Pd, Na, Al, MgO, GaN, AlN, LiF, BN, C, Si, Ge, SiC and 8x8x8 for the remaining. The Hartree-Fock calculations were converged separately with much higher \textit{k}-point samplings. For the CO@Pt(111) system we used 6x6x1 \textit{k}-points for the correlation energies and 16x16x1 \textit{k}-points for the Hartree-Fock energy. For graphene@Ni(111) we used 16x16x1 \textit{k}-points for both Hartree-Fock and correlation. All calculations were performed with a Fermi-smearing of 0.01 eV.

The calculations were performed within the projector-augmented wave framework. For the cohesive energy of solids we found that it is important to include semi-core states of Ag, Pd, Rh and Mg. More details on this point can be found in Ref. [24].

\newpage
\section{Binding energies}
\begin{table}[H!]
\begin{center}
\begin{tabular}{c|c|c|c|c|c|c|c|c}
  & PBE & RPA@LDA & LDA & RPA@PBE & ALDA & rALDA & rAPBE & Exp. \\
	\hline
H$_2$ & 105 & 109 & 113 & 109 & 110 & 111 & 108 & 109 \\
N$_2$ & 244 & 224 & 268 & 225 & 229 & 231 & 228 & 228 \\
P$_2$ & 121 & 116 & 143 & 119 & 118 & 120 & 118 & 117 \\
O$_2$ & 144 & 112 & 174 & 103 & 155 & 118 & 122	& 120 \\
F$_2$ & 53  &  30 &  78 &  24 &  74 &  37 &  41	&  38 \\
Cl$_2$& 65  &  49 &  81 &  40 &  -- &  43 &  59 &  58 \\
CO    & 269 & 242 & 299 & 234 & 287 & 256 & 260	& 259 \\
HF    & 142 & 130 & 161 & 127 & 157 & 139 & 140	& 141 \\
H$_2$O& 234 & 222 & 264 & 218 & 249 & 229 & 229	& 233 \\
CO$_2$& 416 & 360 & 472 & 348 &  -- & 381 & 383	& 389 \\ 
C$_2$H$_2$&415&383& 460 & 374 & 421 & 406 & 404	& 405 \\
C$_2$H$_4$&571&537& 632 & 528 & 578 & 562 & 559	& 564 \\
CH$_4$& 420 & 404 & 462 & 400 & 426 & 420 & 416	& 419 \\
NH$_3$& 302 & 290 & 337 & 291 & 296 & 297 & 293	& 297 \\
\hline
MAE & 0.41 & 0.52 & 1.76 & 0.75 & 0.53 & 0.15 & 0.10 & \\
MAPE& 7.4  & 6.3  & 24   & 10.2 & 11.7  & 3.1  & 1.47 &
\end{tabular}
\end{center}
\caption{Atomization energies of small molecules. The ALDA results are taken from Ref. [15]. Experimental results are corrected for zero point vibrational energies, non-adiabatic coupling and relativistic effects [21]. Numbers are in kcal/mol except the bottom line which shows the mean absolute percentage error.}
\label{tab:molecule}
\end{table}

\begin{table}[H!]
\begin{center}
\begin{tabular}{c|c|c|c|c|c|c|c}
    & PBE  & RPA@PBE & LDA  & RPA@LDA  & rALDA& rAPBE&  Exp \\ 
\hline
C   & 7.73 & 6.99 & 8.94 & 6.83 & 7.54 & 7.58 & 7.55 \\
SiC & 6.38 & 5.96 & 7.38 & 5.89 & 6.44 & 6.50 & 6.48 \\
Si  & 4.55 & 4.32 & 5.63 & 4.37 & 4.82 & 4.79 & 4.68 \\
Ge  & 3.72 & 3.55 & 4.64 & 3.72 & 3.95 & 3.90 & 3.92 \\
BN  & 6.94 & 6.47 & 8.05 & 6.40 & 6.90 & 6.79 & 6.76 \\
LiF & 4.28 & 4.11 & 4.87 & 3.92 & 4.33 & 4.44 & 4.46 \\
LiCl& 3.36 & 3.31 & 3.80 & 3.20 & 3.38 & 3.56 & 3.59 \\
NaF & 3.81 & 3.67 & 4.32 & 3.44 & 3.67 & 3.97 & 3.97 \\
NaCl& 3.10 & 3.08 & 3.47 & 2.95 & 2.97 & 3.30 & 3.34 \\
AlP & 4.07 & 4.05 & 4.81 & 3.99 & 4.03 & 4.29 & 4.32 \\
AlN & 5.67 & 5.50 & 6.58 & 5.43 & 5.63 & 5.76 & 5.85 \\
GaP & 3.44 & 3.43 & 4.30 & 3.21 & 3.39 & 3.60 & 3.61 \\
GaN & 4.33 & 4.25 & 5.32 & 4.04 & 4.26 & 4.43 & 4.55 \\
MgO & 4.98 & 4.94 & 5.86 & 4.81 & 5.07 & 5.18 & 5.20 \\
Al  & 3.43 & 3.14 & 4.00 & 3.06 & 2.97 & 3.47 & 3.43 \\
Na  & 1.08 & 0.98 & 1.23 & 0.97 & 1.07 & 1.08 & 1.12 \\
Pd  & 3.70 & 3.83 & 5.07 & 3.86 & 3.93 & 3.81 & 3.94 \\
Rh  & 5.71 & 5.50 & 7.53 & 5.30 & 5.73 & 5.79 & 5.78 \\
Cu  & 3.47 & 3.34 & 4.49 & 3.35 & 3.57 & 3.55 & 3.52 \\
Ag  & 2.47 & 2.75 & 3.60 & 2.77 & 2.91 & 2.79 & 2.98 \\ 
\hline
MAE & 0.18 & 0.29 & 0.74 & 0.38 & 0.16 & 0.05 & \\
MAPE& 4.4  & 6.9  & 16.2 & 8.7  & 4.1  & 1.37  &
\end{tabular}
\end{center}
\caption{Cohesive energies of solids per atom. Experimental results are corrected for zero point vibrational energies [11]. Numbers are in eV except the bottom line which shows the mean absolute percentage error.}
\label{tab:solid}
\end{table}

\section{Lattice constants}
\begin{table}[H!]
\begin{center}
\begin{tabular}{c|c|c|c|c}
  & PBE & RPA@PBE & rAPBE & Exp. \\
	\hline
N$_2$ & 1.102 & 1.111 & 1.101 & 1.098 \\
F$_2$ & 1.415 & 1.447 & 1.435 & 1.412
\end{tabular}
\end{center}
\caption{Bond lengths of the N$_2$ and F$_2$ molecules calculated with PBE, RPA and rAPBE. Numbers are in {\AA}.}
\label{tab:molecule}
\end{table}

\begin{table}[H!]
\begin{center}
\begin{tabular}{c|c|c|c|c|c|c|c}
    & PBE  & RPA@PBE & rAPBE&  Exp \\ 
\hline
C   & 3.57 & 3.58 & 3.57 & 3.553 \\
Si  & 5.48 & 5.45 & 5.45 & 5.421 \\
Na  & 4.20 & 4.28 & 4.29 & 4.214 \\
Pd  & 3.94 & 3.93 & 3.95 & 3.876
\end{tabular}
\end{center}
\caption{Lattice constants of four solids calculated with PBE RPA and rAPBE. Experimental results are corrected for zero point anharmonic effects [11]. Numbers are in {\AA}.}
\label{tab:solid}
\end{table}

\end{document}